\documentclass{PoS}
\pdfoutput=1

\usepackage{graphics}
\usepackage{graphicx}
\usepackage{amsmath}
\usepackage{bbm}
\usepackage{cite}
\newcommand{\hs}[1]{\hspace*{#1 pt}}
\newcommand{\vs}[1]{\vspace*{#1 pt}}

\def\be{\begin{equation}}
\def\ee{\end{equation}}
\def\bea{\begin{eqnarray}}
\def\eea{\end{eqnarray}}
\def\eps{\epsilon}
\def\nnb{\nonumber}
\def\ep{\epsilon}
\def\eps{\epsilon}
\def\dps{\displaystyle}

\def\zc{z_f}
\def\ub{\bar u}

\title{Non-leptonic $B$-decays at two-loops in QCD}

\ShortTitle{Non-leptonic $B$-decays at two-loops in QCD}

\author{\speaker{Tobias Huber}\\
        Theoretische Physik 1, Naturwissenschaftlich-Technische Fakult\"at,
 Universit\"at Siegen, Walter-Flex-Stra{\ss}e 3, D-57068 Siegen, Germany\\
        E-mail: \email{huber@tp1.physik.uni-siegen.de}}

\abstract{I review the status of the calculation of two-loop QCD corrections to non-leptonic $B$-decays in the framework of QCD factorisation. In the case of heavy-to-light decays I will 
cover the leading penguin amplitudes $a_4^u$ and $a_4^c$. For heavy-to-heavy transitions I will discuss the amplitude $a_1(D^{(*)}\pi)$. In both cases I will present some computational details, especially on how to obtain analytical results for the master integrals in a canonical basis, which in turn enables us to derive analytical expressions for almost all terms in the amplitudes.}

\FullConference{ Loops and Legs in Quantum Field Theory - LL 2014,\\
		 27 April - 2 May 2014 \\
		 Weimar, Germany }

\begin{document}

\section{Introduction}

Non-leptonic exclusive decays of $B_{(s)}$ and $D_{(s)}$ mesons offer a rich and interesting phenomenology and play a decisive role in quantifying the amount of CP violation, the most subtle phenomenon of flavour physics. The main focus is on two-body decays and observables like branching fractions, (CP)-asymmetries, and polarisations; whereas three-body decays also allow for more detailed decay characteristics such as Dalitz-plot distributions. The wealth of experimental data from
flavour-factories (BaBar, Belle, CLEO, BES~III etc.), hadron colliders (Tevatron and LHC), and in the future also from ``superflavour''-factories (Belle~II), will yield ever more precise measurements of numerous observables from more than a hundred different final states of non-leptonic decays.

The high precision of experimental data clearly justifies every effort to obtain accurate
theoretical predictions for nonleptonic $B$- and $D$-decays. This task is -- however -- complicated by the purely hadronic initial and final state, where QCD effects from many different scales arise. In the last decades there have been several attempts to solve this problem. Early approaches, nowadays called the naive factorization, express the hadronic matrix elements as products of a decay constant times a form factor, see e.g.\ \cite{Bauer:1986bm}. During the last 15 years, several refinements have been developed. The most successful ones are based on flavour symmetries such as isospin, U-spin or flavour-SU(3) (see e.g.~\cite{Jung:2009pb}); and/or factorization, like pQCD~\cite{Keum:2000ph} or QCD factorisation~\cite{Beneke:1999br,Beneke:2000ry}.

Precise theoretical descriptions in non-leptonic heavy-meson decays, together with large data sets from experiments, are indispensable ingredients to sharpen our understanding of the strong dynamics at scales up to $\sim 5$~GeV. On the other hand, the plethora of observables, together with possible correlations among them, makes hadronic decays a viable tool for indirect searches for new physics~(NP). Despite the fact that they are not as sensitive to new phenomena compared to rare (semi-)leptonic or radiative transitions, new interactions may manifest themselves dominantly in purely hadronic transitions, especially if they violate the CP symmetry.

\section{Theoretical Framework}

The decays of heavy quarks are described in an effective five-flavour theory where the top quark and the heavy gauge bosons $W^\pm$,~$Z$ are integrated out. The resulting effective weak Hamiltonian reads~\cite{Buchalla:1995vs,Chetyrkin:1997gb}
\begin{align}
\dps \mathcal{H}_{eff} =& -\frac{4\,G_F}{\sqrt{2}} \sum_{p=u,c} \lambda_p \left[ C_1 Q_1^p
+ C_2 Q_2^p + \sum_{k=3}^6 C_k Q_k + C_8 Q_8\right] + \rm{h.c.} \, ,
\end{align}
with $\lambda_p = V_{pb}V^*_{pd}$ and
\begin{align}
Q_{1,2}^p =& (\bar d_L \gamma^\mu \{T^a,\mathbbm{1}\} p_L) (\bar p_L \gamma_\mu \{T^a,\mathbbm{1}\} b_L) \, , &&
Q_{3,4}  =& (\bar d_L \gamma^\mu \{\mathbbm{1},T^a\} b_L) \sum_q (\bar q \gamma_\mu \{\mathbbm{1},T^a\} q) \, , \nnb \\
Q_{5,6}  =& (\bar d_L \gamma^\mu\gamma^\nu\gamma^\rho \{\mathbbm{1},T^a\} b_L) \sum_q (\bar q \gamma_\mu\gamma_\nu\gamma_\rho \{\mathbbm{1},T^a\} q) \, , &&
Q_{8\;\;\;\,}=& -\dps\frac{g_s}{16\pi^2} \, m_b \; \bar d_L \; 
\sigma_{\mu\nu} G^{\mu\nu} b_R \; . \hs{42} \nnb
\end{align}

In this article we focus on the QCD factorisation~(QCDF)-approach to non-leptonic $B$-decays~\cite{Beneke:1999br,Beneke:2000ry}. It is a model-independent framework that systematically disentangles short-distance (perturbative) from long-distance (non-perturbative) effects in the heavy-mass limit. The factorisation formula reads
\begin{align}
\dps \langle M_1 M_2 | Q_i | \bar{B} \rangle =& m_B^2 \; F_+^{B \to
M_1}(0) \; f_{M_2}
\int_0^1 du \; \; T_{i}^I(u) \; \phi_{M_2}(u)\nnb \\
\dps +& f_{B} \, f_{M_1} \, f_{M_2} \int_0^\infty d\omega \int_0^1  dv du \; \;
T_{i}^{II}(\omega,v,u) \; \phi_B(\omega)  \; \phi_{M_1}(v) \;
\phi_{M_2}(u) \; .  \label{eq:QCDF}
\end{align}
The quantities $T_{i}^{I,II}$ are the perturbatively calculable hard-scattering kernels, where $T_{i}^{I}$ includes the so-called vertex-corrections and starts at ${\cal O}(1)$, and $T_{i}^{II}$ comprises the contributions from spectator scattering and starts at ${\cal O}(\alpha_s)$. Each of the hard-scattering kernels further splits up into a contribution from ``tree'' and ``penguin'' topologies. Since for $B \to D\pi$ there are no contributions from penguin topologies, and spectator scattering is power-suppressed~\cite{Beneke:2000ry}, we have only the first line of~(\ref{eq:QCDF}) in this case.
The non-perturbative quantities are the transition form factor $F_+^{B \to M_1}$, the decay constants $f_j$, and the distribution amplitudes $\phi_k$ of the heavy and light mesons. Hence, in order to achieve precision predictions one needs both, higher-order calculations for the perturbative quantities, and precise input of non-perturbative parameters. The latter can be obtained for instance from lattice or sum rule calculations.

The QCDF formula~(\ref{eq:QCDF}) is valid to all orders in $\alpha_s$ and to leading order in $\Lambda_{\rm{QCD}}/m_b$. Moreover, the leading term turns out to be real.
Strong phases are thus either induced by perturbative contributions to the hard-scattering kernels
or by power-suppressed terms. Consequently, they are predicted to be parametrically of order ${\cal O}(\alpha_s)$ or ${\cal O}(\Lambda_{\rm{QCD}}/m_b)$.

\section{Motivation for NNLO}

The matrix elements of the effective weak Hamiltonian between an initial and final state can be written as a linear combination of different topological amplitudes. For instance, one finds
\begin{align}
\sqrt{2} \; \langle \pi^- \pi^0 | \, \mathcal{H}_{eff} \, | B^- \rangle \;
   & = \; A_{\pi\pi} \; \lambda_u   \big[\alpha_1(\pi\pi) + \alpha_2(\pi\pi) \big] \, ,\nnb \\
 - \; \langle \pi^0 \pi^0 | \, \mathcal{H}_{eff} \, | \bar{B}^0 \rangle \;
   & = \; A_{\pi\pi} \; \big\{ \lambda_u \big[\alpha_2(\pi\pi) - \alpha_4^u(\pi\pi)\big] - \lambda_c  \, \alpha_4^c(\pi\pi) \big\} \, ,\nnb \\
  \langle \pi^-\bar K^0 | \, \mathcal{H}_{eff} \, | B^- \rangle \;
   &\dps= \; A_{\pi\bar K}\, \left[\lambda^{(s)}_u \, \alpha_4^u + \lambda^{(s)}_c \, \alpha_4^c \right]\, .
\end{align}
Here, $\alpha_1$ and $\alpha_2$ are the colour-allowed and colour-suppressed topological tree amplitudes, respectively, whereas $\alpha_4^{u,c}$ are the leading penguin amplitudes. One clearly sees the different dependences of the various channels on these amplitudes. The decay $B^- \to \pi^-\pi^0$ does not depend on the penguin amplitudes, while $B^- \to \bar K^0\pi^-$ is a pure penguin decay. The decay $\bar B^0 \to \pi^0\pi^0$ has no colour-allowed tree amplitude, and its phenomenology gives rise to some puzzles, e.g.\ the theory prediction of the branching ratio is much smaller than the expermental value, although the error bars are still quite large. A systematic study of this type, together with the NLO predictions for the amplitudes and a thorough phenomenological analysis was carried out in~\cite{Beneke:2003zv}. The analysis at NLO reveals that there is multiple motivation to go to NNLO.
\begin{itemize}
\item For the colour-suppressed tree amplitude $\alpha_2$ there is a large cancellation between LO and NLO, which makes it particularly sensitive to NNLO.
\item As mentioned above, the direct CP-asymmetries are first generated at ${\cal O}(\alpha_s)$. Therefore NNLO is only the first perturbative correction to these quantities and thus crucial for the reduction of scale dependence.
\item From a rather conceptual point of view it is interesting to verify explicitly that factorisation holds to NNLO.
\end{itemize}
Part of the NNLO calculation has already been carried out, namely the one-loop ${\cal O}(\alpha_s^2)$
correction to the hard spectator-scattering~\cite{Beneke:2005vv,Beneke:2006mk,Kivel:2006xc,Pilipp:2007mg,Jain:2007dy}, as well as the two-loop ${\cal O}(\alpha_s^2)$ correction to the tree topology of the vertex kernel~\cite{Bell:2007tv,Bell:2009nk,Beneke:2009ek,Bell:2009fm}. The tree amplitudes $\alpha_1$ and $\alpha_2$ are therefore known to NNLO, and the result could be achieved completely analytically. A numerical investigation of the NNLO tree amplitudes shows that the two-loop correction is quite sizable both in the vertex and spectator-scattering part, with a large cancellation between the two.

The only NNLO-piece which is still missing at leading power is the vertex-correction to the penguin amplitudes. Including all known pieces the latter read
\allowdisplaybreaks{
\begin{align}
 \alpha_4^u(\pi \pi) =& -0.029 - [0.002 + 0.001i ]_V + [0.003- 0.013i ]_P+ [ ?? \;
 +\; ?? \; i]_{{\cal O}(\alpha_s^2)} \nnb \\
+& \left[ \frac{r_{\rm sp}}{0.485} \right]
   \big\{ [0.001]_{\rm LO} + [0.001 +0.000i]_{HV+HP} + [0.001]_{\rm tw3} \big\} \nnb \\
=& -0.024^{+0.004}_{-0.002} + (-0.012^{+0.003}_{-0.002})i   \, ,\\ 
 \alpha_4^c(\pi \pi) =& -0.029 - [0.002 +  0.001i ]_V - [0.001+ 0.007i ]_P 
 + [ ?? \; + \;  ??  \; i]_{{\cal O}(\alpha_s^2)}   \nnb \\
+ & \left[ \frac{r_{\rm sp}}{0.485} \right]
   \big\{ [0.001]_{\rm LO} + [0.001 +0.001i]_{HV+HP} + [0.001]_{\rm tw3} \big\} \nnb \\
   =& -0.028^{+0.005}_{-0.003} + (-0.006^{+0.003}_{-0.002})i \, ,
\end{align}}

\noindent and the question marks stand for the aforementioned yet unknown corrections, which are subject to our current investigations.

\section{Details of the Calculation and Results}

\FIGURE[h]{
\includegraphics[width=0.35\textwidth]{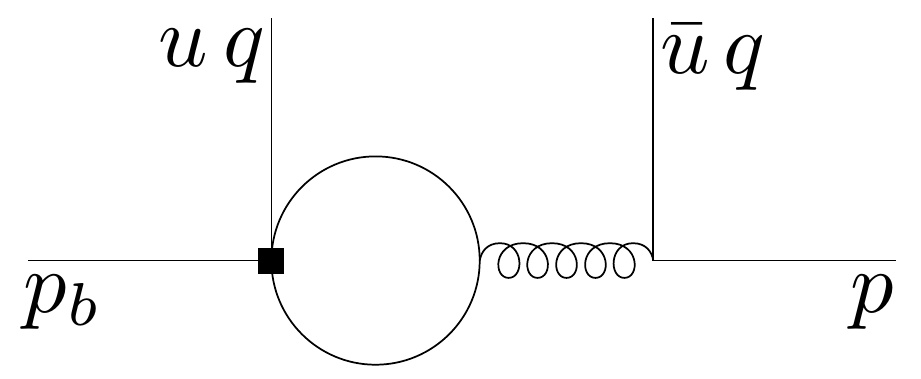}
\caption{Kinematics of the heavy-to-light decay. External lines are on-shell with $p_b^2=m_b^2$ and $p^2=q^2=0$. The fermion-loop can have mass $m=0, \, m_c, \, m_b$. The black square denotes a vertex from an operator in the effective weak Hamiltonian.}\label{fig:kinematics}}

The calculation of the NNLO correction to the vertex-kernel of the leading penguin amplitudes $\alpha_4^u$ and $\alpha_4^c$ amounts to the evaluation of $\sim70$ Feynman diagrams, a subset of them is shown in Fig.~\ref{fig:diagrams}. The one-loop ${\cal O}(\alpha_s^2)$ contribution of the chromomagnetic dipole operator $Q_8$, depicted in the right panel in Fig.~\ref{fig:diagrams}, was calculated in~\cite{Kim:2011jm}. All other contributions are genuine two-loop diagrams.

\FIGURE[t]{
\hs{-3}\includegraphics[scale=.35]{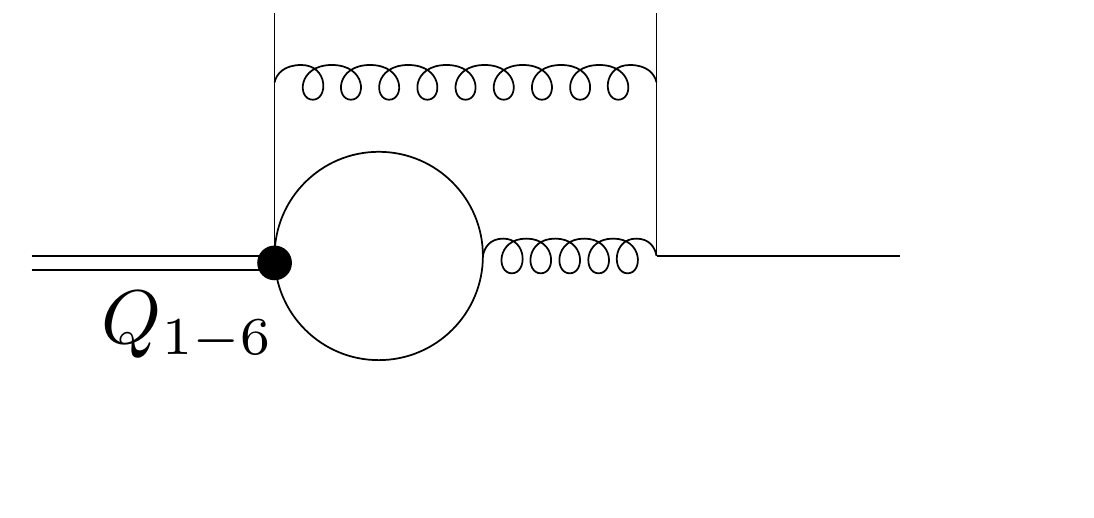}
 \hs{-10}\includegraphics[scale=.35]{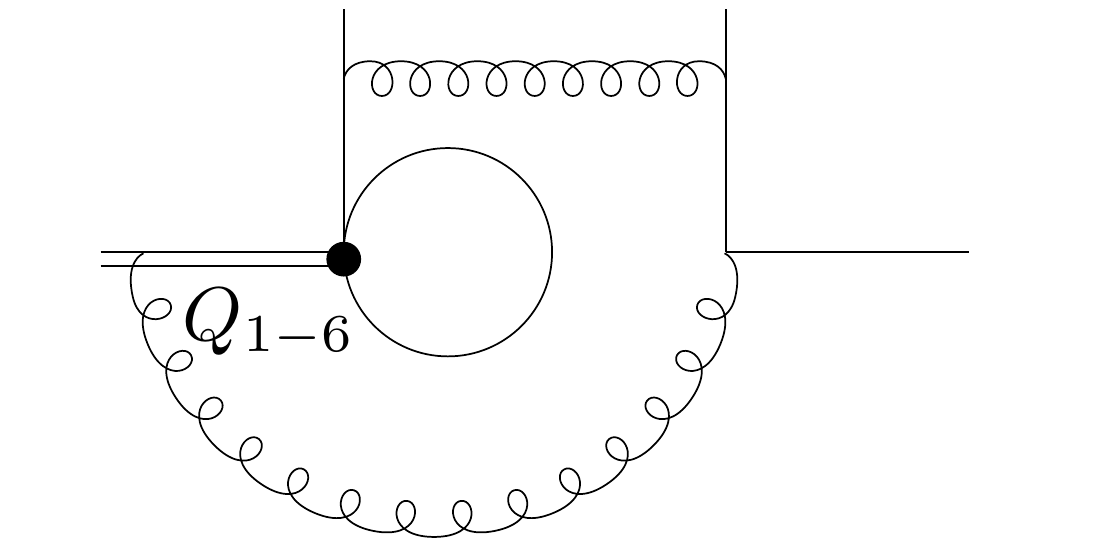}
 \hs{-0}\includegraphics[scale=.35]{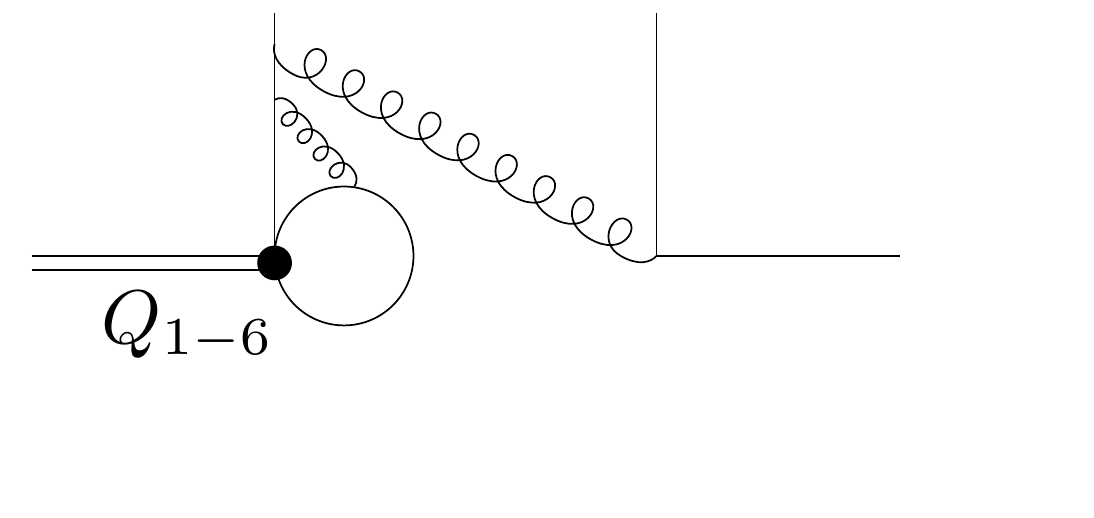}
 \hs{-10}\includegraphics[scale=.35]{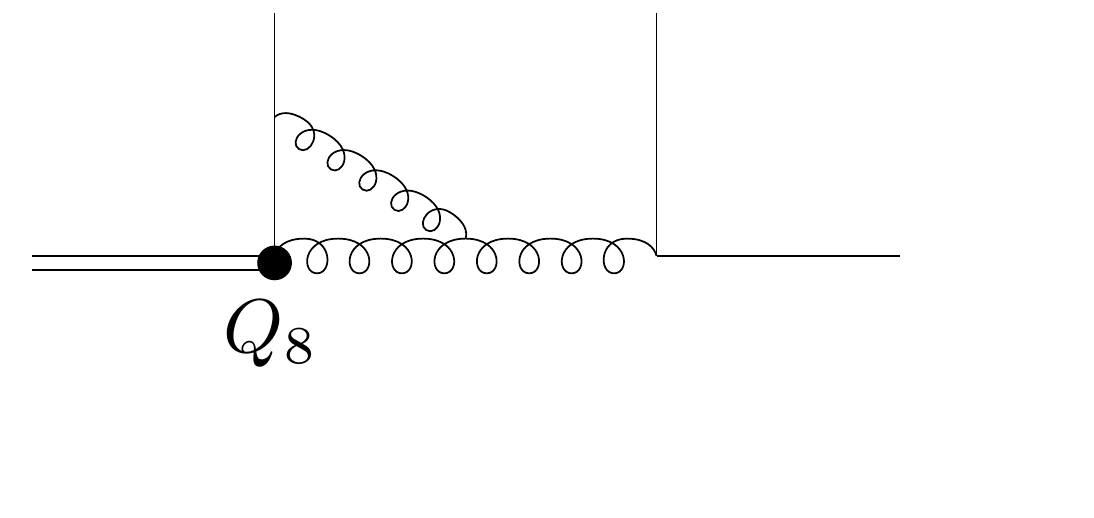}

\vs{-15}

\caption{Sample diagrams at NNLO. Diagrams like the first one in which a gluon directly connects the internal fermion loop to the fermion line on the right are referred to as ``genuine'' penguin diagrams, whereas diagrams like the second and the third are representatives of so-called ``exotic'' diagrams. The diagram on the right denotes a one-loop ${\cal O}(\alpha_s^2)$ contribution from the chromomagnetic dipole operator $Q_8$.}\label{fig:diagrams}}

\FIGURE[b]{
\includegraphics[width=0.95\textwidth]{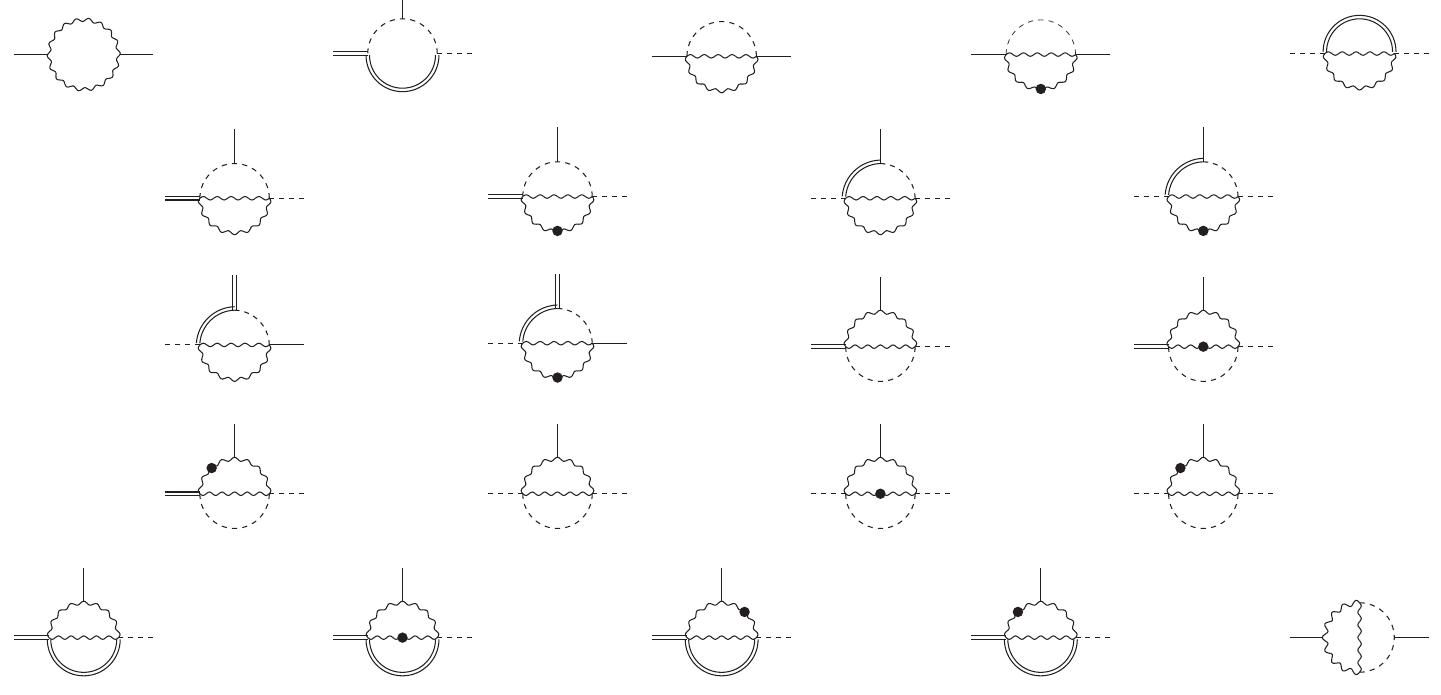}
\caption{Master integrals from genuine penguin diagrams at two loops. Dashed/wavy/double internal lines denote propagators with mass $0/\sqrt{\zc} m_b/m_b$ (where $\zc=m_f^2/m_b^2$). Dashed/solid/double external lines correspond to virtualities $0/\ub m_b^2/m_b^2$. Dotted propagators are taken to be squared.}\label{fig:masters}}

The kinematics of the process is depicted in Fig.~\ref{fig:kinematics}. The problem depends on two dimensionless variables, the momentum fraction $\ub=1-u$, and the mass ratio $\zc \equiv m_f^2/m_b^2$, with $f=c,b$. The reduction of the amplitude is done by techniques which by now have become standard in multi-loop computations. We use dimensional regularisation with $D=4-2\ep$, reduce the tensor structure via Passarino-Veltman relations, followed by reduction of the scalar integrals to master integrals using AIR~\cite{Anastasiou:2004vj}, FIRE~\cite{Smirnov:2008iw}, and an in-house routine. This procedure results in 29 yet unknown master integrals. The ones that result from the diagram class in the left panel of Fig.~\ref{fig:diagrams} are shown in Fig.~\ref{fig:masters}. The techniques used to solve the master integrals are based on the expansion of hypergeometric functions~\cite{Huber:2005yg,Huber:2007dx}, Mellin-Barnes representations~\cite{Czakon:2005rk}, sector decomposition~\cite{Borowka:2012yc}, and differential equations~\cite{Kotikov:1990kg}. Especially the latter method in a canonical basis, proposed in~\cite{Henn:2013pwa}, where the system of differential equations assumes the form
\begin{align}
d \; \vec f = \ep \; \tilde A \; \vec f \; ,
\end{align}
proves to be a powerful method to obtain analytical results for the hard scattering kernel, which are suitable for the convolution with the light-cone distribution amplitude (LCDA). In the following we give two examples of master integrals in a canonical basis.

Our first example is the topology that consists of the first four master integrals in the last line of Fig.~\ref{fig:masters}. Their canonical basis is shown in Fig.~\ref{fig:canonicalI}. The topology has to be enlarged by lower-line integrals to make the system of differential equations complete, but in order to keep the formulas short we omit these lower-line integrals here. The differential equations are conveniently written in terms of the variables
$r \equiv  \sqrt{1-4\zc}$ and $s \equiv  \sqrt{1-4\zc/\ub}$,
which render the pre-factors that accompany the masters rational. We find, for instance
\FIGURE[t]{
\hs{-3}\includegraphics[width=1.08\textwidth]{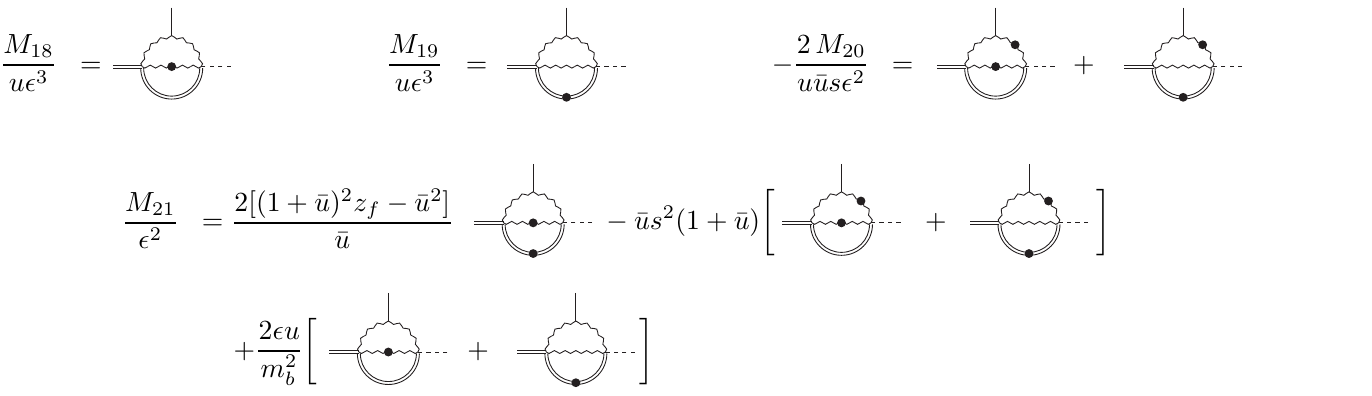}

\vs{-10}

\caption{Master integrals in the canonical basis I. The lines have the same meaning as in Fig.~\ref{fig:masters}.}\label{fig:canonicalI}}
\begin{align}
\dps \frac{\partial M_{19}}{\partial s} \; \; =&\frac{4 \eps  M_{18} \, r \, \left(r^2+1\right)}{(r^2+1)^2-4s^2}-\frac{2 \eps  M_{19} \, r \, \left(r^2+s^2-2\right)}{(1-r^2)
   (r^2-s^2)}+\frac{4 \eps  M_{20} \, r \, s}{\textstyle(r^2+1)^2-4 s^2}
   -\frac{\eps  M_{21} \, r \, \left(r^2+1\right)}{(r^2+1)^2-4 s^2} \, .
\end{align}
The structure of this formula reveals that the solution can be written as an iterated integral over rational weight functions. Together with the boundary conditions that the integrals vanish either in $s=r$ ($M_{18,19}$) or $s=+i\infty$ ($M_{20,21}$), this completely fixes the solution. The entire alphabet of rational weight functions for our iterated integrals reads
\begin{equation}
\dps \left\{ s \; , \; 1\pm s \; , \; r \; , \; 1\pm r \; , \; r\pm s \; , \;  r^2+1 \dps \pm 2s \; , \; 1+2\textstyle\sqrt{\zc}\dps \pm s \; , \;  1-2\textstyle\sqrt{\zc}\dps \pm s\right\} \, .
\end{equation}
The complete set of master integrals in a canonical basis, together with all boundary conditions and analytical solutions, can be found in~\cite{Bell:2014zya}.
\FIGURE[b]{
\hs{-3}\includegraphics[width=1.21\textwidth]{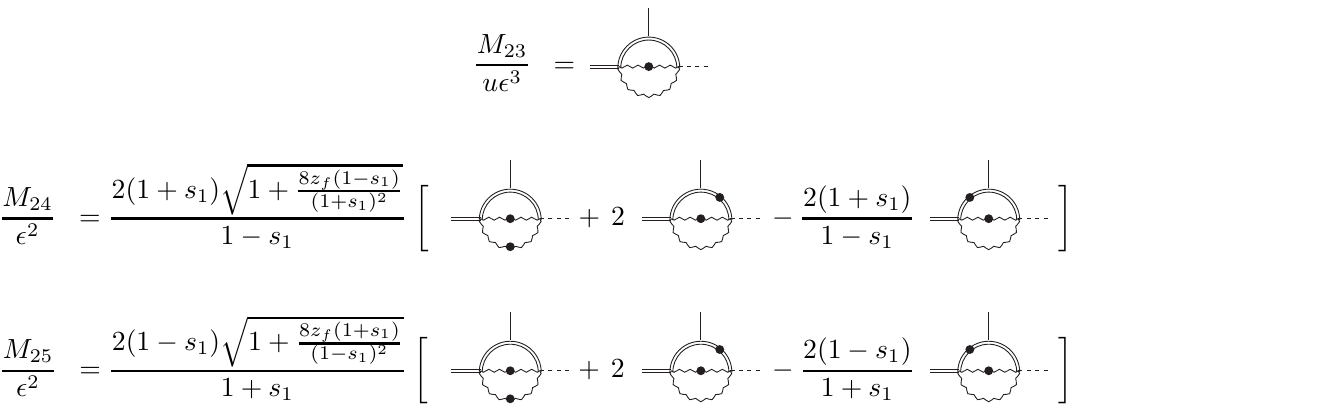}

\vs{-10}

\caption{Master integrals in the canonical basis II. The lines have the same meaning as in Fig.~\ref{fig:masters}.}\label{fig:canonicalII}}

Our second example consists of a topology of three master integrals that stem from diagrams of the ``exotic'' type, see Fig.~\ref{fig:diagrams}. Their canonical basis is given in Fig.~\ref{fig:canonicalII}, where we again omit lower-line integrals of this topology. It turns out that it is convenient to trade the variable $\ub$ for $s_1 \equiv  \sqrt{1-4/\ub}$, which yields differential equations with irrational factors,
\begin{align}
\dps \frac{\partial M_{23} }{\partial s_1} \; \; = \frac{2\,\eps\,  M_{23}  \,s_1 \! \left( 5 - s_1^2 \right)}
   {\left( 1 - s_1^2 \right) \,\left( 3 + s_1^2 \right) } - 
  \frac{\eps\, M_{24}  \,\left( 3 - s_1 \right) }
   {4 (1-s_1^2) \, 
     {\textstyle{\sqrt{1+\frac{8\,\zc (1-s_1)}{(1 + s_1)^2}}}}\dps} + 
  \frac{\eps\, M_{25}  \,\left( 3 + s_1 \right) }
   {4 (1-s_1^2) \, 
     {\textstyle{\sqrt{1+\frac{8\,\zc (1 + s_1)}{(1-s_1)^2}}}}\dps} \, .
\end{align}
It is far from obvious that there exists a variable transformation that rationalises the pre-factors, and that the solution falls into the class of iterated integrals. In~\cite{Bell:2014zya}, we present such a transformation and give further conditions that enable us to obtain the solution to these master integrals in a completely analytical form.

The results of the master integrals enable us to compute the NNLO correction to the vertex-kernel of the penguin amplitudes. We have the result for $\alpha_4^u$ completely analytically, and $\alpha_4^c$ as an accurate interpolation in $\zc$. Here the numerical evaluation of iterated integrals by means of GiNaC~\cite{Vollinga:2004sn} is very useful. Both results will be presented in~\cite{Bell:2014B}.

\section{The decay $\bar B \to D\pi$}

\FIGURE[t]{
\includegraphics[width=0.23\textwidth]{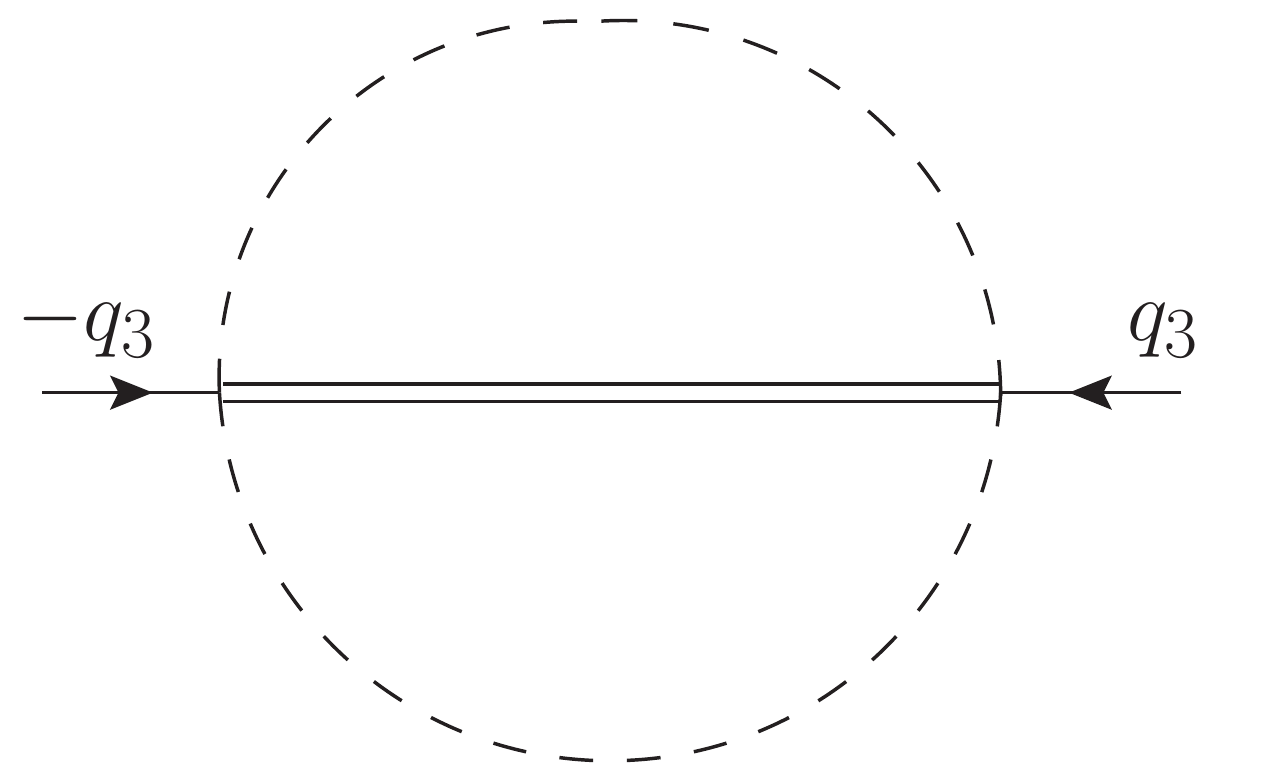}\hs{6}
\includegraphics[width=0.23\textwidth]{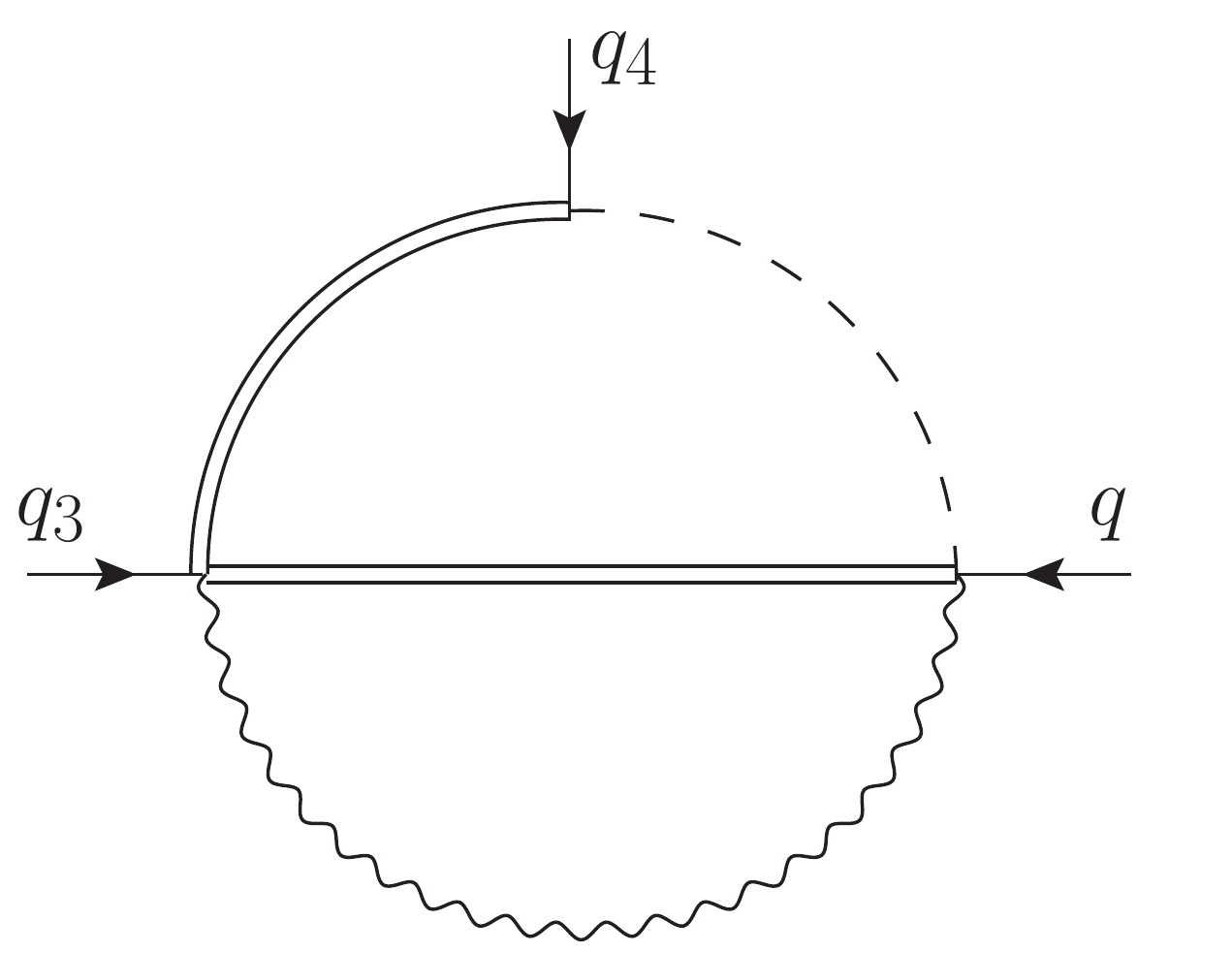}\hs{6}
\includegraphics[width=0.23\textwidth]{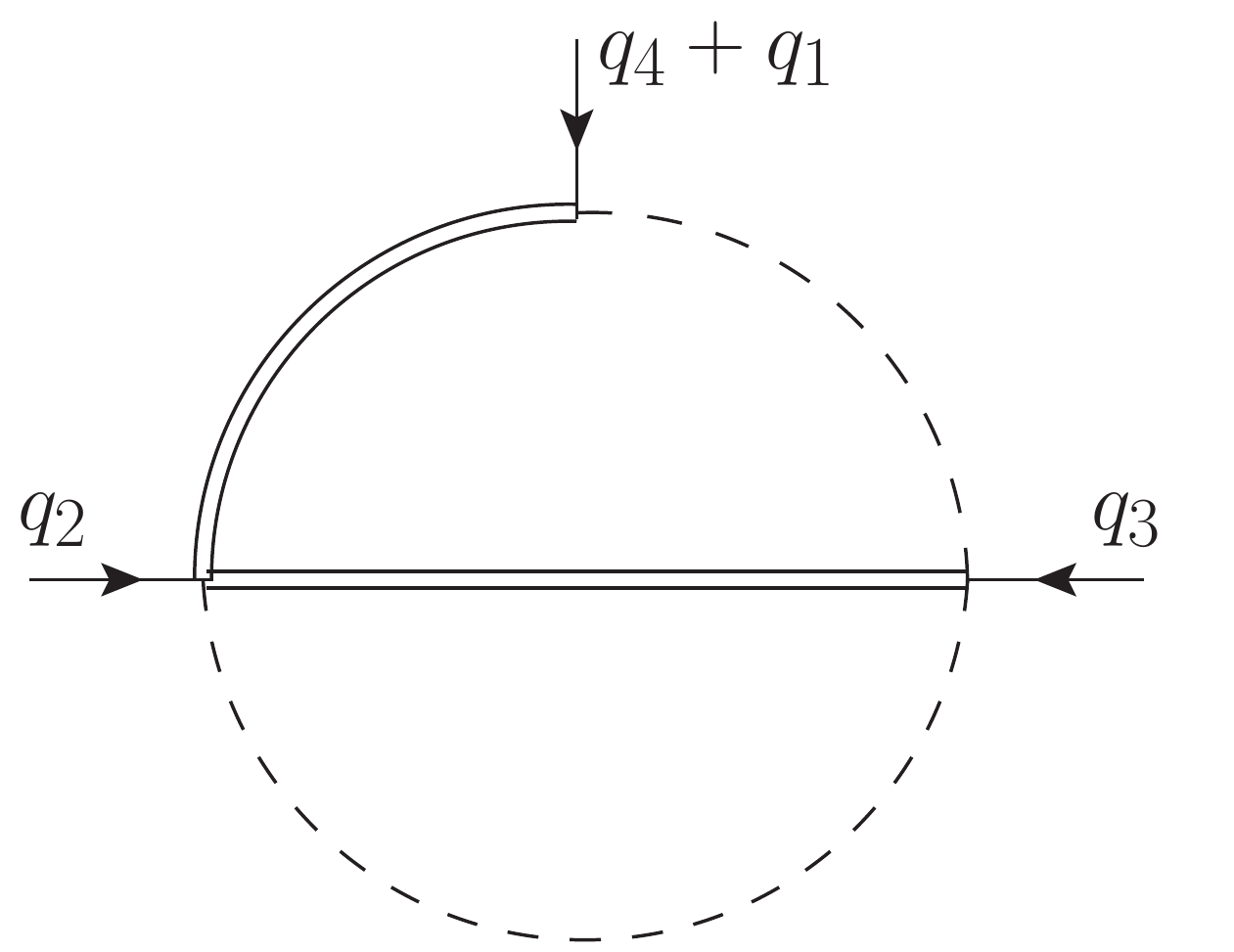}\hs{6}
\includegraphics[width=0.23\textwidth]{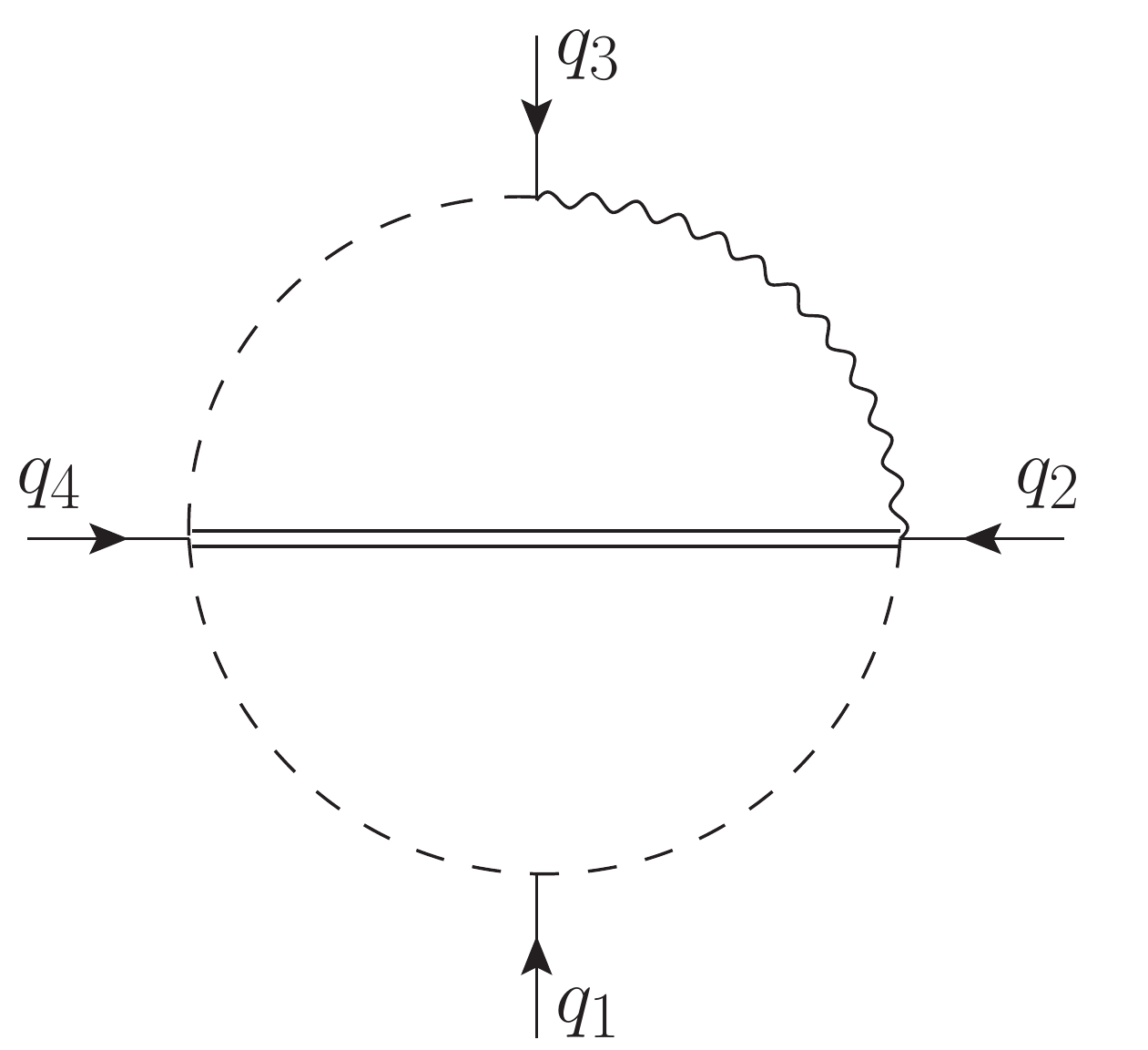}
\caption{Selected master integrals for the decay $\bar B \to D\pi$. The dashed, the curly and the double lines represent quarks of mass $0, \, m_c, \, m_b$, respectively. 
  The external momenta satisfy $q_1^2=q_2^2=0$, $q_3^2= m_c^2$, and $q_4^2= m_b^2$.}\label{fig:mastersBDpi}}
In the case when a final state flavour is charm, we have only the colour-allowed tree-amplitude. In  QCDF the branching ratio is given by~\cite{Beneke:2000ry}
\begin{align}
  \Gamma(\bar{B}^0 \rightarrow  D^{+} \pi^{-})= \frac{G_F^2(m_B^2-m_D^2)^2 |\vec{q}|}{16 \pi m_B^2} |V^*_{ud}	 V_{cb}|^2  |a_1(D \pi)|^2 	\, f^2_\pi \,F^2_0(m^2_\pi)  \, ,
 \end{align}
The results for $a_1$ to NLO accuracy are given below for a light meson $L$~\cite{Beneke:2000ry},
 \begin{align}
| a_1(\bar{B}^0 \rightarrow D L)|=&(1.055^{+0.019}_{-0.013})  -(0.013^{+0.011}_{-0.006}) \alpha^{L}_{1} \, , \nonumber \\
| a_1(\bar{B}^0 \rightarrow D^* L)|=&(1.054^{+0.018}_{-0.017})  -(0.015^{+0.013}_{-0.007}) \alpha^{L}_{1} \, . \label{a1NOL}
 \end{align}
In case of $\pi$ and $\rho$ we have $\alpha^{\pi(\rho)}_{1}=0$ and for the kaon $|\alpha^{K}_{1}|<1$ is assumed. Also in this decay we have sensitivity to NNLO because the NLO QCDF corrections to $a_1$ are small since they are colour-suppressed and accompanied by small Wilson coefficients.

The calculation again amounts to $\sim70$ Feynman diagrams which are shown in Figs.~15 and~16 of~\cite{Beneke:2000ry}. They also result in about two dozens of yet unknown master integrals, which also depend on two scales $u$ and $\zc$. A sample of the masters is shown in Fig.~\ref{fig:mastersBDpi}. We completed all masters, and the hard-scattering kernel of the colour-singlet operator $Q_2$~\cite{Huber:2014kaa}. For the convolution with the LCDA a canonical basis will be most desirable.

\section*{Acknowledgments}

I would like to thank the organisers of ``Loops and Legs 2014'' for creating a pleasant and inspiring atmosphere. I would like to thank Guido Bell, Martin Beneke, Susanne Kr\"ankl, and Xin-Qiang Li for collaboration on the topics covered in this article. This work is supported by DFG research unit FOR 1873 ``Quark Flavour Physics and Effective Field Theories''.

\end{document}